\documentclass{article}
\usepackage{url} %this package should fix any errors with URLs in refs.
\usepackage{graphicx}
\usepackage{soul}

\begin{document}

\title{SIR-HUXt - a particle filter data assimilation scheme for assimilating CME time-elongation profiles.}

\author{Luke Barnard, Mathew Owens, Chris Scott, Matthew Lang, Mike Lockwood\\
Department of Meteorology, University of Reading, UK}

\maketitle

\section*{Abstract}
We present the development of SIR-HUXt, the integration of a sequential importance resampling (SIR) data assimilation scheme with the HUXt solar wind model. SIR-HUXt is designed to assimilate the time-elongation profiles of CME fronts in the low heliosphere, such as those typically extracted from heliospheric imager data returned by the STEREO, Parker Solar Probe, and Solar Orbiter missions.

We use Observing System Simulation Experiments (OSSEs) to explore the performance of SIR-HUXt for a simple synthetic CME scenario of a fully Earth directed CME flowing through a uniform ambient solar wind, where the CME is initialised with the average observed CME speed and width. These experiments are performed for a range of observer locations, from $20^{\circ}$ to $90^{\circ}$ behind Earth, spanning the L5 point where ESA's future Vigil space weather monitor will return heliospheric imager data for operational space weather forecasting.

We show that, for this CME scenario, SIR-HUXt performs well at constraining the CME speed, and has some success at constraining the CME longitude. The CME width is largely unconstrained by the SIR-HUXt assimilations, and further experiments are required to determine if this is related to the specific CME scenario, or is a more general feature of assimilating time-elongation profiles. An analysis of rank-histograms suggests the SIR-HUXt ensembles are well calibrated, with no clear indications of bias or under/over dispersion. Improved constraints on the initial CME speed lead directly to improvements in the CME transit time to Earth and arrival speed.
For an observer in the L5 region, SIR-HUXt returned a $69\%$ reduction in the CME transit time uncertainty, and a $63\%$ reduction in the arrival speed uncertainty. This suggests SIR-HUXt has potential to improve the real-world representivity of HUXt simulations, and therefore has potential to reduce the uncertainty of CME arrival time hindcasts and forecasts.

\section{Introduction}

Coronal Mass Ejections (CMEs) are large eruptions of magnetised plasma from the Sun's atmosphere. CMEs are the primary cause of severe and extreme space weather at Earth, and so understanding the heliospheric evolution of CMEs, and forecasting their arrival at Earth receives a significant amount of research effort. Currently, numerical simulations of the real-world heliospheric evolution of coronal mass ejections (CMEs) have significant uncertainty. This is partly evidenced by the large average errors in CME arrival time of $\pm10~h$ \cite{riley2018}. These uncertainties are caused by a range of factors, but appear to be dominated by uncertainties in the initial and boundary conditions of the solar wind models, with uncertainty in the ambient solar wind structure and CME parameters introducing similar levels of uncertainty \cite{riley2021}. These uncertainties therefore limit our ability to draw scientific conclusions from the simulations of real-world CMEs, and limit the skill of space weather forecasts of CMEs.

This motivates the pursuit of methods with which to reduce the uncertainty on numerical simulations of real-world CMEs. Data Assimilation (DA) methods have excellent potential for improving the representivity of solar wind numerical models. The objective of DA is to combine information from simulations and observations to provide an optimal estimate of the state of a dynamical system. Heliospheric DA is still a relatively new research topic, but progress is beginning to be made.

\cite{lang2017} explored how the Local Ensemble Transform Kalman filter could be used to assimilate in-situ observations of solar wind plasma properties into the ENLIL magnetohydrodynamic (MHD) solar wind model, which demonstrated clear improvements in the representivity of the ENLIL simulations. The Buger Radial Variational Data Assimilation (BRaVDA) scheme was developed in \cite{lang2019}, in which a variational DA scheme was coupled to the hydrodynamic (HD) HUX solar wind model \cite{riley2011}, for the assimilation of observations of the solar wind speed. Experiments with synthetic observations and solar wind speed observations from the STEREO spacecraft showed that BRaVDA reduced the errors in the solar wind speed predictions at Earth. This work was extended by \cite{lang2021} to the HUXt model, a HD solar wind model with explicit time-dependence \cite{owens2020a,barnard2022}, in which it was shown that over the period 2007-2014, BRaVDA returned a $31\%$ reduction in the RMSE of hindcasts of the solar wind speed at Earth.

These works have so far focused on the assimilation of in-situ observations of solar wind plasma properties, but progress has also been made on the assimilation of remote sensing observations, such as those provided by heliospheric imagers (HIs) \cite{eyles2008,howard2008} and interplanetary scintillation (IPS) \cite{fallows2022}. For example, \cite{barnard2020} showed that an ensemble of solar-wind-CME simulations with the HUXt model could be weighted by the time-elongation profiles of CMEs derived from the STEREO Heliospheric Imager (HI) data. This weighting prioritised ensemble members that more closely matched the observed time-elongation profile, and led to up to $20\%$ improvements in hindcasts of the CMEs arrival time at Earth. Similarly, \cite{iwai2021} demonstrated how assimilating Interplanetary Scintillation (IPS) observations of 12 halo CMEs into the SUSANOO-CME MHD model led to improvements in the predicated Earth arrival times of these CMEs.

Although \cite{barnard2020} demonstrated that HI data contains useful information on CMEs that can be used to constrain the HUXt solar wind simulations, they did not use formal DA methods. In this work, we present the development of SIR-HUXt, which couples a sequential importance resampling (SIR) particle filter DA scheme with the HUXt solar wind model. SIR-HUXt is constructed to assimilate time-elongation profiles of a CMEs flank, such as those typically extracted from the STEREO-HI data \cite{Davies2009,barnard2015a,barnard2017}. This is an important milestone towards the development of DA schemes that can directly assimilate the HI intensity data into solar wind numerical models. We present a first test of SIR-HUXt by using Observing System Simulation Experiments (OSSEs) to investigate the performance of SIR-HUXt for a simple synthetic CME scenario and for a range of observer locations relative to Earth. 

This article proceeds with section \ref{sec:methods} describing the models and methods we use, including the HUXt numerical model, the background to the SIR algorithm, and on OSSEs. Section \ref{sec:results} presents the results of the OSSEs, and our conclusions are presented in section \ref{sec:conclusions}.

\section{Methods and Data}
\label{sec:methods}

\subsection{HUXt}
\label{sec:huxt}
HUXt is an open source numerical model of the solar wind, developed in Python \cite{owens2020a,barnard2022}. It is a 1D radial model that uses a reduced-physics approach to produce solar wind simulations that emulate the solar wind flows produced by 3-D MHD models, but at a small fraction of the computational cost.

The motivation for developing HUXt is that the models simplicity and computational expense permits the development of certain experiments and techniques that would typically be too expensive with 3-D MHD models. For example, the particle filter data assimilation experiments in this study require $\approx 10^6$ 5-day simulations of the inner heliosphere, which is currently an impractical demand of 3-D MHD solar wind models with widely available computing resources. 

In this work, HUXt is run with its default configuration. The radial grid spans $30~R_{\odot}$ to $240~R_{\odot}$, with a grid step of $1.5~R_{\odot}$. The time-step is $8.7~$minutes. There are 128 evenly space longitudinal bins, although to save on computation, and as we are only examining Earth-directed CMEs, the simulation domain only spans the longitude range of $\pm~70^{\circ}$. 

CMEs are included in HUXt via the Cone CME parameterisation, in which CMEs are represented as a time-dependent velocity perturbation to the model inner boundary. Six parameters are required to specify the initiation of a Cone CME; the initiation time; the speed; the angular width; the source longitude and latitude; and the radial thickness of the perturbation. Further details of the Cone CME parameterisation in HUXt are given in \cite{owens2020a,barnard2021a,barnard2022}. CMEs are tracked through HUXt simulations by inserting test particles into the flow on the CME surface at the model inner boundary. These test particles then passively advect with the flow and are followed at all time steps out to the model’s outer boundary.

Pseudo-observers are used with HUXt to compute the time-elongation profile of the Cone CME flank, to emulate the time-elongation data products typically derived from Heliospheric Imager observations e.g. \cite{Davies2009,barnard2015a,barnard2017,pant2016}. This is achieved by computing the elongation of each particle on the CME boundary and finding the particle with maximum elongation in an observer’s field of view. A better solution would be to forward model the observations from Heliospehric Imager instruments by performing Thomson scattering simulations with HUXt output. However, the HUXt equations are derived from incompressible hydrodynamics, and so only the flow speed is solved for, not the flow mass density. This prohibits a fully self-consistent forward modelling of Heliospheric Imager data from HUXt simulations. Consequently, tracking the maximum elongation of the CME tracer particles is a necessary approximation. However, both \cite{barnard2020} and \cite{chi2021} showed that this approach returned time-elongation profiles that compared favourably to those extracted directly from STEREO-HI images, which gives us confidence this approximation is reasonable.

\subsection{Sequential Importance Resampling (SIR)}
\label{sec:sir}
The objective of data assimilation is to provide an optimal estimate of the state of a system by combining the information from both a model and observations of the system, taking proper account of the uncertainties on each. 

This can be expressed mathematically via Bayes' theorem, which states that,

\begin{equation}
    p(\psi |\theta) = \frac{p(\theta|\psi)p(\psi)}{p(\theta)}.
\end{equation}

The factors in this equation are typically separated into into several colloquially named terms. The "prior", $p(\psi)$, is the probability density of the model being in a specific state, in the absence of any other external information. The "likelihood", $p(\theta|\psi)$, which is the probability density of obtaining a set of observations $\theta$, given a model state $\psi$. The "evidence", $p(\theta)$, is the probability density of obtaining a set of observations although in most practical examples the evidence becomes a normalising constant that can be ignored. Finally, the "posterior", $p(\psi |\theta)$, is the conditional distribution of model states given a set of observations.

Computation of the posterior, or approximations to it, is the focus of data assimilation. The posterior provides the optimal estimate of the state of the system, representing the distribution of model states that are most consistent with the observations. In practical geophysical examples, it is not possible to fully characterise the posterior distribution, and different data assimilation methodologies are used to infer certain properties of the posterior e.g. its mean, mode, or variance \cite{ledimet1986,burgers1998}. A particle filter is set of a data assimilation methodologies that aims to approximate the full posterior distribution via an ensemble of "particles" \cite{vanleeuwen2009,chorin2009,ades2013,browne2015,fearnhead2017,potthast2019}.

Sequential Importance Resampling (SIR) is a method of particle filtering that can be used for sequential data assimilation \cite{vanleeuwen2009, fearnhead2017}. In SIR, the posterior is approximated by the analysis of an ensemble of simulations, or "particles". The prior is approximated by generating an ensemble of simulations that reflects the uncertainty in the models initial and boundary conditions. The model evolves the ensemble forward in time, until a set of observations are available. At the observation time, an analysis is performed which weights each simulation in accordance with its agreement with the observations. Then, this weighted ensemble is used to generate a new ensemble of simulations which are closer to the observations. The model then resumes advancing the simulations forward in time, until the next set of observations are available. The data assimilation proceeds in this way, performing sequential analysis steps when observations are available. The posterior distribution, at some specific time, is approximated by the distribution of the ensemble after an analysis step.

In this work we develop SIR-HUXt, a coupling of an SIR scheme with the HUXt solar wind model, with the objective of assimilating the time-elongation profiles of CME fronts, such as those that can be derived from white light heliospheric imaging. SIR-HUXt essentially functions as a form of parameter estimation, returning estimates of the posterior of the Cone CME parameters that are most consistent with the observed the time-elongation profile of a CME. The following subsections describe the specifics of the SIR algorithm used in SIR-HUXt.

\subsubsection{Initial Ensemble Generation}
\label{sec:ensemble_generation}
The initial ensemble is generated by perturbing a subset of the Cone CME parameters only, following a similar method to \cite{barnard2020}. Specifically, perturbations are applied to the Cone CME speed, angular width, and longitude. We focus on these three parameters only as they are probably the most important parameters for determining if and when a CME impacts Earth \cite{pizzo2015,riley2021}, whilst considering all the Cone CME parameters would be too computationally expensive for this proof-of-concept study.

The random perturbations for each parameter are drawn from a uniform distribution that represents the observational uncertainty on that parameter, and the perturbation is added to the best-guess of the true Cone CME parameter. For the speed, width, and longitude, the spread of the uniform perturbation distributions is $\pm 10\%$, $\pm 5^{\circ}$, and $\pm 5^{\circ}$, respectively.

The true uncertainty distributions are unlikely to be uniform, but there is not yet good knowledge on what form the observational uncertainties take. And so, in the absence of better knowledge, we follow \cite{barnard2020} and use the uniform distribution.

The size of the ensemble must be large enough to avoid "filter degeneracy", where the ensemble essentially collapses into one particle. This occurs when one particle has much larger weight than others in the ensemble such that it dominates the resampling procedure, leading to new particles that are degenerate. The upper bound on the sample size is determined by the availability of computational resources. Here we use an ensemble size of 50. This was determined empirically during our initial experiments. Future work should look to optimise the ensemble size, but for our purposes 50 members appears to perform sufficiently well. 

\subsubsection{Particle weighting}
During the analysis phase of an SIR scheme, a weight must be assigned to each particle by comparison with the available observations. This requires computing an approximation to $p(\theta |\psi)$, the likelihood of recording an observation $\theta$, given the modelled state $\psi$. 

With SIR-HUXt, we are investigating the usefulness of assimilating the time-elongation profile of a CME flank which could be observed by Heliospheric Imager-like instruments. Therefore, in this context, we must compute the likelihood of an observed CME flank elongation value for a specific modelled flank elongation. 

Computation of the time-elongation profiles of the Cone CMEs is described in section \ref{sec:huxt}, whilst generation of the pseudo-observations, where Gaussian noise is added to the time-elongation profiles, is described in section \ref{sec:osse}

To compare the simulated and observed flank elongation, we use an assumed Gaussian likelihood profile for $p(\theta |\psi)$, which is centred on the simulated flank elongation with a spread of $0.15^{\circ}$. As far as we are aware, there is no good \emph{a priori} or empirical knowledge of how this likelihood profile should be structured. However, we believe that a Gaussian is a reasonable approximation, which can be refined in future. 

Then, after computing the observation likelihood $l_i$ for each ensemble member, its weight $w_i$ is computed as the normalised likelihood over the $N$ ensemble members,

\begin{equation}
w_{i} = \frac{l_{i}}{\sum_{i=1}^{N}l_{i}}.
\end{equation}

These weights are then used in the resampling procedure.

\subsubsection{Resampling}

Kernel density estimation is used to compute the resampling of the ensemble members. Each Cone CMEs state is represented by its speed, width, and longitude. To resample the ensemble, we must draw samples from the kernel density estimate of the joint-distribution of these 3 parameters. However, scale separation of the parameters, particularly the CME speed from the width and longitude, means it is necessary to rescale the parameters before computing the kernel density estimate. Therefore, we compute the z-scores of each parameter before computing the kernel density estimate of the joint distribution and resampling, using a Gaussian kernel with a bandwidth of 0.2. This bandwidth value was arrived at experimentally. If the bandwidth is too large, the resampled ensemble will not be drawn towards the observations and so the assimilation achieves little to nothing. Whereas if it is too small, the resampled ensemble can be pulled too aggressively towards a highly weighted particle, making filter degeneracy more probable. 

\begin{figure}
\includegraphics[width=\textwidth]{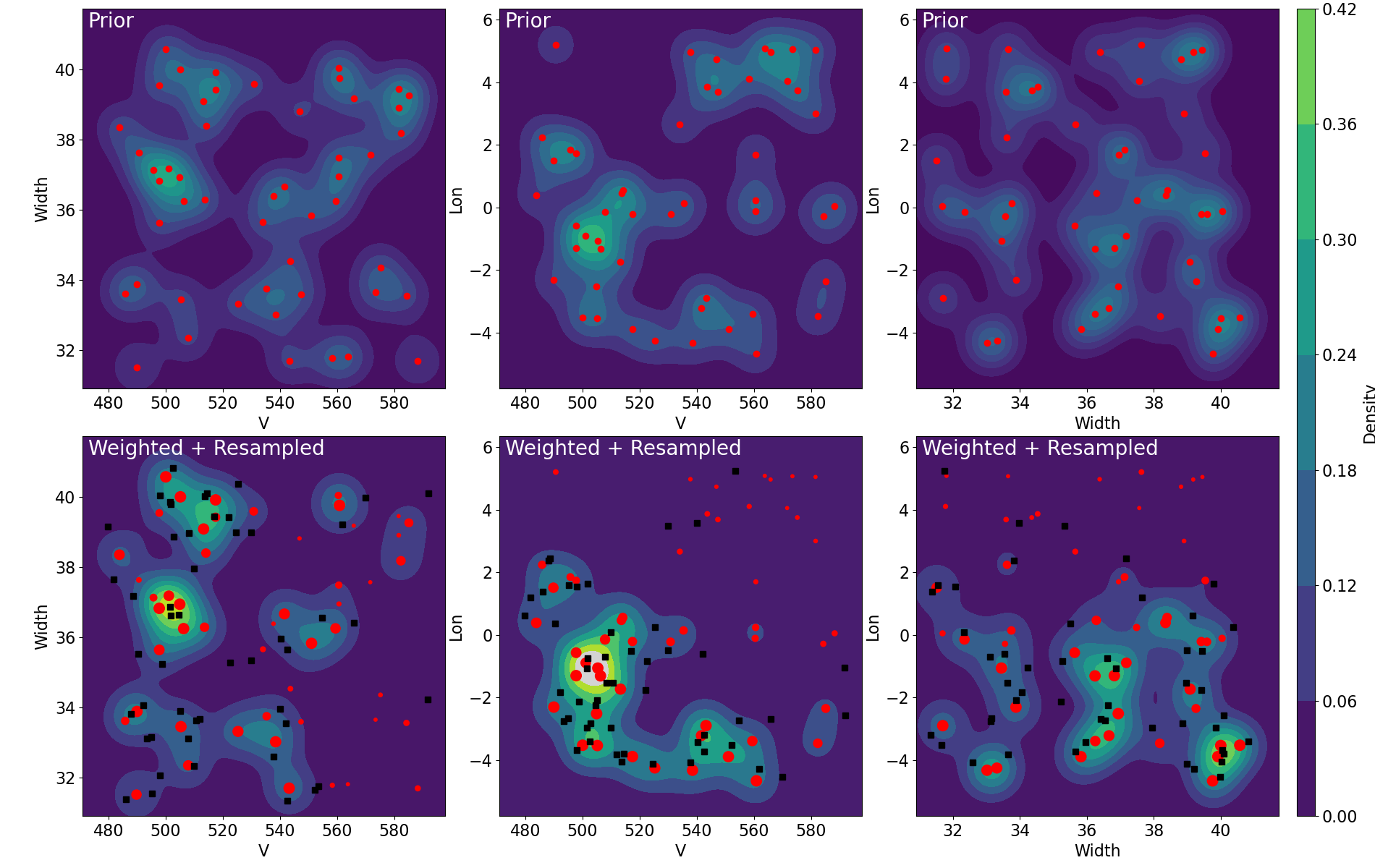}
\caption{The top row shows the Cone CME parameters of the initial ensemble (prior) in red dots, with contours of the kernel density estimates of their distribution. The bottom row shows the Cone CME parameters of the initial ensemble (prior) in red dots, with a size proportional to their weight determined in the SIR analysis step. The contours show the kernel density estimates of these weighted distributions, whilst the black dots show the resampled Cone CME parameters that form the ensemble that will be advanced to the next analysis step.}
\label{fig:resample_example}
\end{figure}

Figure \ref{fig:resample_example} shows how the resampling procedure works in practice. The top row shows the Cone CME parameters of the initial ensemble (the prior) as red dots, for each pairing of the Cone CME speed, width, and longitude. The kernel density estimate of the distribution of these points is also contoured. These distributions are relatively uniform, given the sample size of 50, as would be expected from the uniform perturbation functions that generate the prior. In the bottom row the red dots show the same prior Cone CME parameters, but with a size proportional to their weight determined in the first analysis step of an SIR computation. Here, the contours instead show the distribution of the weighted prior Cone CME parameters. The black squares show the resampled Cone CME parameters that form the new ensemble that will be advanced to the next analysis step. It is clear that the new ensemble is closer to the prior Cone CME parameters that had larger weights, but is not overly concentrated around particles with the largest weights. 

\subsection{Observing System Simulation Experiment (OSSE)}
\label{sec:osse}
An Observing System Simulation Experiment (OSSE) is a method with which we can assess the potential benefits of integrating a data assimilation scheme into a physical model of a system \cite{zeng2020a}. OSSEs are controlled experiments using simulations of synthetic scenarios that allow us to explore the usefulness of different observation networks and/or data assimilation schemes \cite{cucurull2021}.

These experiments begin by using a model to simulate a "ground truth". Observations of this ground truth are generated by combining a forward model that emulates the observations from the ground truth with realistic observational noise. Then, these emulated observations are assimilated into the data assimilation scheme, where the physical model is initialised with perturbed initial and/or boundary conditions relative to the "ground truth" simulation. Through this process we can assess the ability of a data assimilation scheme to recover the "ground truth".

Here we use OSSEs configured as a "twin experiment", where we perform the same experiment with both HUXt and SIR-HUXt, to assess the performance of the SIR scheme relative to an ensemble of HUXt simulations without data assimilation. This is the same general method as that employed by \cite{lang2017}, who investigated the use the Local Ensemble Transform Kalman Filter with the WSA-ENLIL solar wind model, and in \cite{lang2019}, in the development of the Burger Radial Variational Data Assimilation with the HUX solar wind model.

\begin{figure}
\includegraphics[width=\textwidth]{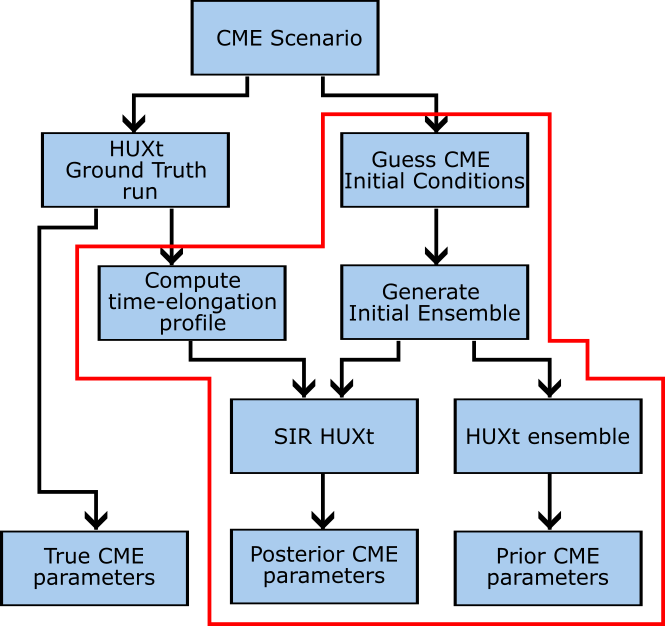}
\caption{A flow chart showing the configuration of the OSSE experiments. The stages encapsulated in the red box are repeated for 100 iterations, with each iteration using a different random realisation of the guessed CME initial conditions, the computed time-elongation profile, and the initial ensemble.}
\label{fig:osse_flow}
\end{figure}

Figure \ref{fig:osse_flow} present a flow diagram of the configuration of the OSSE experiments. To collect statistics on the performance of the SIR-HUXt scheme for a particular combination of CME scenario and Observer, the steps bounded in red are repeated 100 times, with different random realisations of; the guess at the CME initial conditions; the generation of the initial ensemble; and the observed time-elongation profile of the CME flank. The following subsections describe the individual steps in this flow chart.

\subsubsection{CME scenario}
\label{sec:cme_scenario}
\begin{figure}
\includegraphics[width=\textwidth]{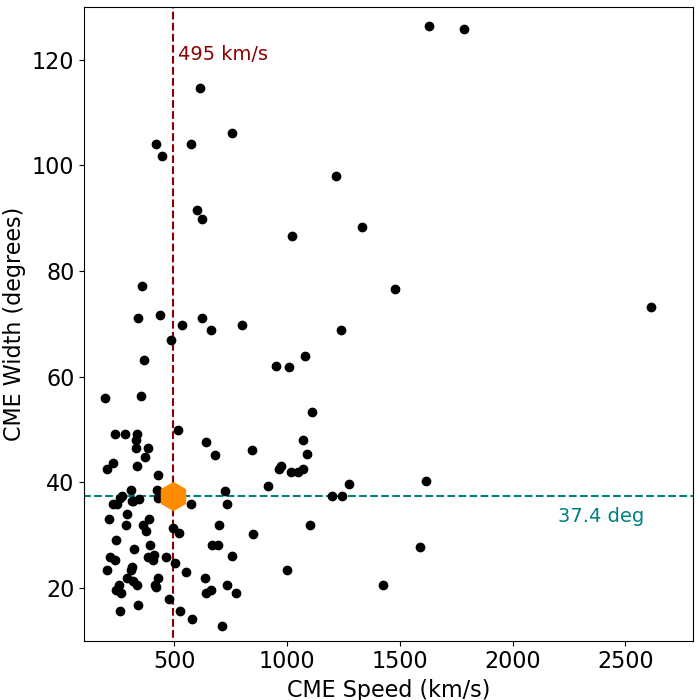}
\caption{The black dots mark the CME speeds and widths from the HELCATS CME classifications determined from stereoscopic fits to coronagraph observations. The orange hexagon marks the speed and width parameters of the Cone CME scenario used in this study, which correspond to the median of the speed and width distributions, which are marked with dashed and dotted lines.}
\label{fig:helcats_cme_scenario}
\end{figure}

We use a Cone CME scenario to develop the SIR-HUXt system with. This scenario reflects the climatological average CME. To build this scenario, we analysed the distribution of observed CME speeds and widths provided by KINCAT database in the HELCATS project. The KINCAT data are described in \cite{barnes2020} and \cite{pluta2019}, and consist of graduated cylindrical shell (GCS) fits \cite{thernisien2011} of 122 CMEs observed in the STEREO COR2 coronagraphs \cite{howard2008}. These GCS fits return estimates of the CME apex speed and the angular half-width. These data are presented as a scatter plot in Figure \ref{fig:helcats_cme_scenario}. 

We compute the medians of the CME speed and (full) width to define the average CME scenario. The Cone CME is fully Earth-directed, having the same source longitude and latitude as Earth, and is initialised 1 hour after the model start time. These values are summarised in table \ref{tab:cme_scenario} and shown by the orange hexagon in Figure \ref{fig:helcats_cme_scenario}.

\begin{table}
\caption{Cone CME scenario used in the OSSEs }
\label{tab:cme_scenario}
\centering
\begin{tabular}{l c}
\hline
Parameter  & Value  \\
\hline
Speed ($km~s^{-1}$)  & 495 \\
Full Width (deg) & 37.4 \\
Longitude (deg) & Earth's \\
Latitude (deg) & Earth's \\
Thickness ($R_{\odot}$ & 1.0 \\
\hline
\end{tabular}
\end{table}

\begin{figure}
\includegraphics[width=\textwidth]{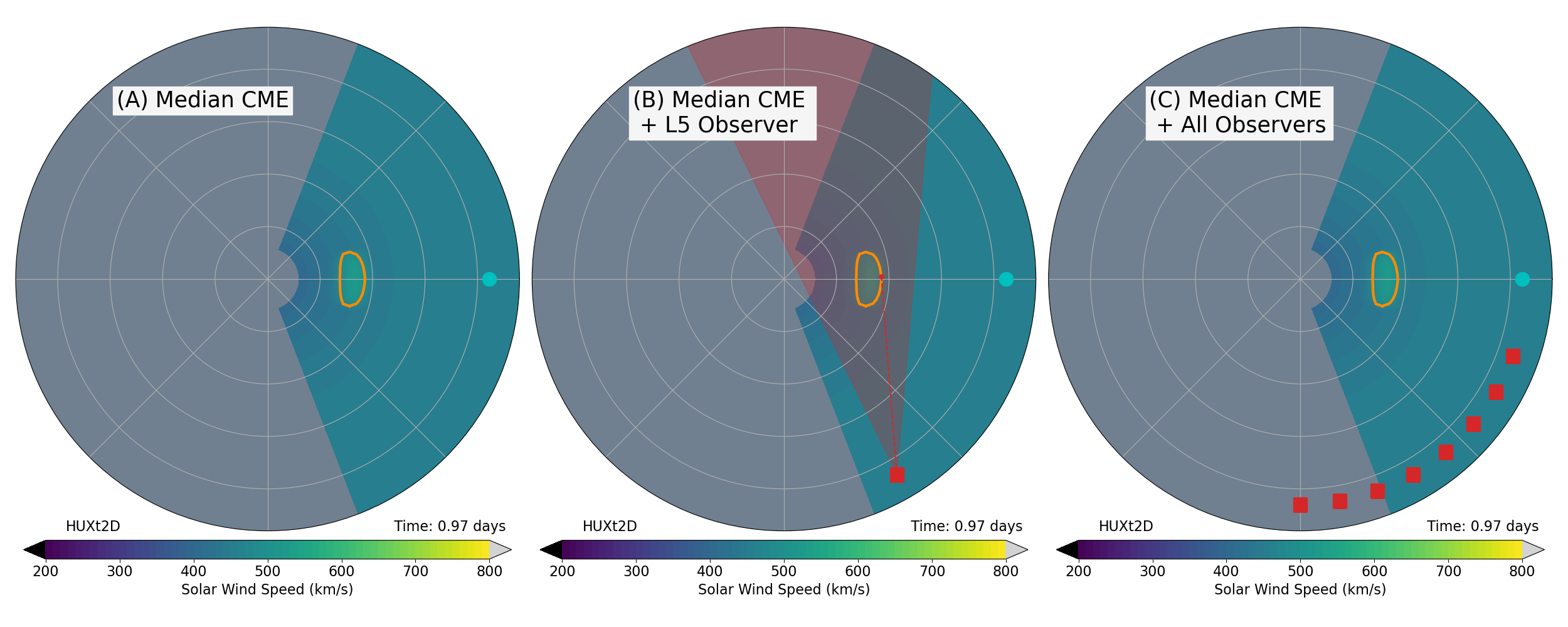}
\caption{These panels show snapshots of the HUXt simulations of the median CME scenario approximately 1 day after model initialisation. The background solar wind speed is uniform, being 400km/s at the model inner boundary. Earth is marked by the cyan dot, whilst the CME boundary is marked by the orange line. Panel B shows the field-of-view (red shaded region) of the L5 pseudo-observer (red square), as well as the tracked flank of the Cone CME (red dashed line). Panel C shows the locations 8 pseudo-observers used in this study.}
\label{fig:huxt_cme_scenario}
\end{figure}

A uniform ambient solar wind is used with the Cone CME scenario, with the ambient solar wind speed at the inner boundary being set to $400 km/s$. We choose to use a uniform ambient solar wind in these experiments to reduce the complexity of the system whilst we develop SIR-HUXt. Future experiments will explore the impact of both different CME scenarios and structured solar wind on the performance of the SIR-HUXt. Figure \ref{fig:huxt_cme_scenario} presents snapshots from the ground-truth simulations of each of the Cone CME scenario. The Cone CME boundaries are marked by the orange lines, whilst Earth is marked by the cyan circle. This scenario provides the ground truth simulation against which the performance of the SIR scheme will be assessed. 

\subsubsection{HUXt Ground Truth Run}

A HUXt simulation is produced using the unperturbed CME scenario, including calculations of the CME transit time to Earth and arrival speed, computed using the standard HUXt tools \cite{owens2022b}

\subsubsection{Compute Time-Elongation Profile}

An observer tracks the time-elongation profile of the Cone CME flank in the HUXt ground truth run, as described in section \ref{sec:huxt}. The flank is tracked over the elongation range spanning $4^{\circ}$ to $35^{\circ}$, with observations recorded every 174 minutes (corresponding to 20 steps of HUXt's native time-step). To these time-elongation profiles, Gaussian noise is added to the elongations with a standard deviation of $0.1^{\circ}$. We consider this a reasonable lower limit on the elongation uncertainty, as analysis of time-elongation profiles extracted from STEREO Heliospehric Imager data suggest that elongation uncertainties of $\approx 0.5^{\circ}$ are typical \cite{williams2009,mostl2011,barnard2015a,barnard2017}.

In each SIR-HUXt OSSE, only observations from one observer are assimilated. However, to investigate the impact of observer longitude, we run the experiments with 8 observers at longitudes spanning $-90^{\circ}$ to $-20^{\circ}$ in steps of $10^{\circ}$, all situated in the same latitudinal plane as Earth. Figure \ref{fig:huxt_cme_scenario} panel B shows the locations of the 8 observers that track the time-elongation profiles of the Cone CMEs. Panel C also shows the field-of-view of the observer (red shaded region) situated at $-60^{\circ}$, corresponding to the L5 location.

\subsubsection{Guess CME Initial Conditions}

In both research and forecasting simulations of real world CMEs, we do not have perfect knowledge of a CMEs initial conditions. These parameters must be estimated from observations and/or empirical relations. To emulate this process here, in each OSSE, we make a guess at the CME initial conditions by applying a perturbation to the ground truth Cone CME parameters of the CME scenario. The perturbations are calculated using the same procedure as is used to generate the initial ensemble of Cone CME parameters, as described in section \ref{sec:ensemble_generation}. To summarise, these are perturbations to the Cone CME speed, width, and longitude, derived from uniform distributions that approximate the uncertainties on the estimated CME parameters.

\subsubsection{Generate Initial Ensemble}

The initial ensemble of Cone CME parameters, which is used in both the SIR-HUXt simulations and the HUXt ensemble, is generated according to the procedure described in \ref{sec:ensemble_generation}. In the generation of the initial ensemble, the guess of the CMEs initial conditions is used as the best estimate to which the perturbations are applied.

\subsubsection{SIR-HUXt}

SIR-HUXt takes the initial ensemble and the observed time-elongation profile and performs eight iterations of the SIR analysis. These eight analysis steps are the maximum that can be performed consistently across all experiments, within the observers constraints of the field-of-view extending to only $35^{\circ}$ elongation, and recording observations every 174 minutes. At each analysis step, the Cone CME parameters of each ensemble member are recorded, as are the CME transit time to Earth and arrival speed, which is computed using the standard HUXt tools.

\subsubsection{HUXt ensemble}

The HUXt ensemble run proceeds by simply generating a HUXt simulation for each member of the initial ensemble. Similarly, each of the Cone CME parameters are recorded, as are the CME transit time to Earth and arrival speed, computed using the standard HUXt tools.

\subsubsection{The true, prior, and posterior CME parameters}

From the above simulations we have knowledge of the true CME parameters, including transit time to Earth and arrival speed, as well as the prior distributions of these parameters, returned by the HUXt ensemble, and the posterior distributions, returned by SIR-HUXt. These data are then used in the statistical assessment of the performance of SIR-HUXt relative to a simple ensemble of HUXt runs. 

\section{Results}
\label{sec:results}

\subsection{An L5 Observer of the median CME scenario}
This Observer-CME scenario combination is highly relevant scenario for future space weather forecasting, for two reasons. Firstly, the median CME scenario reflects the most frequently occurring class of CME. Secondly, as ESA's Vigil mission will provide heliospheric imaging data from L5 for use in operational space weather forecasts, the L5 Observer approximates the time-elongation profiles that might be obtained operationally by Vigil's heliospehric imager.

\subsubsection{Example of one realisation of SIR-HUXt analysis}

\begin{figure}
\includegraphics[width=\textwidth]{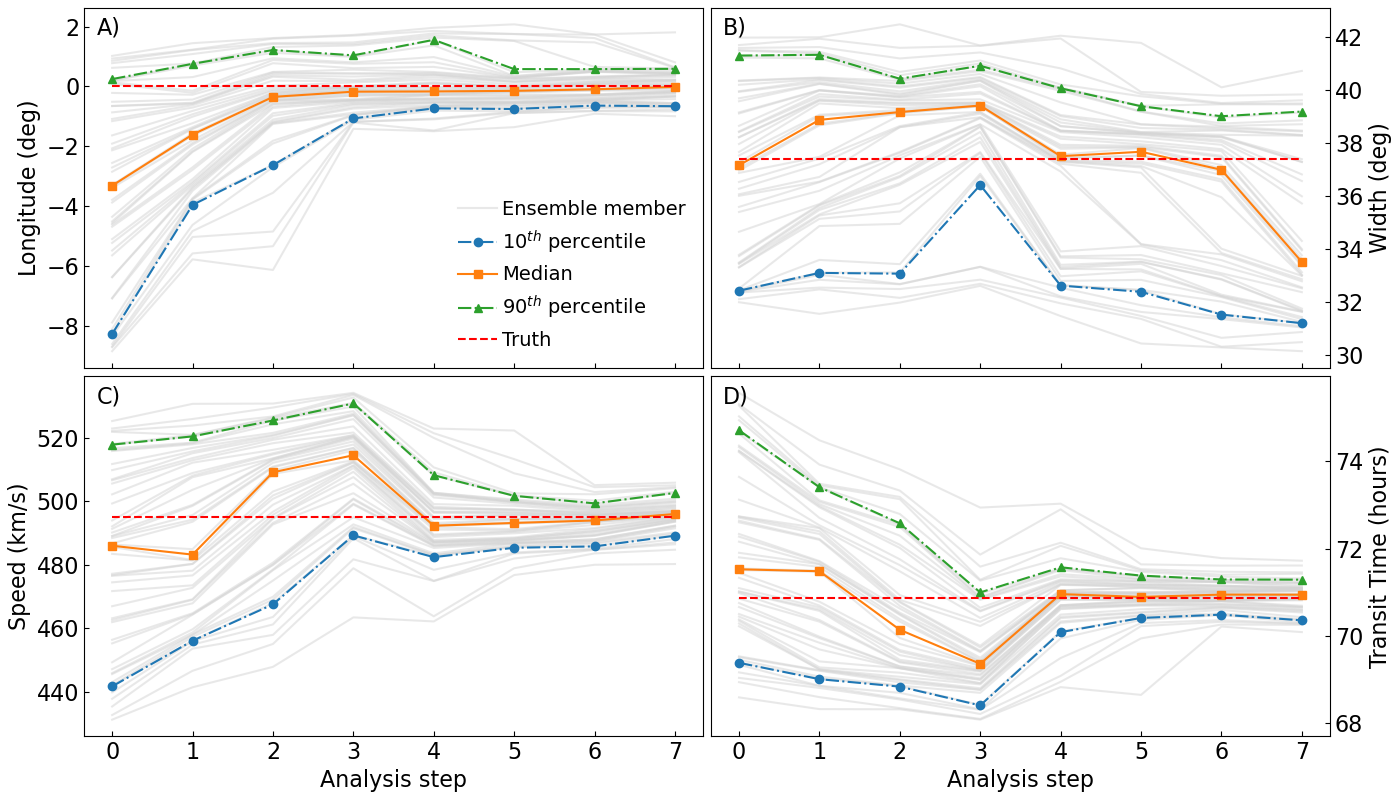}
\caption{These panels show the evolution of the SIR-HUXt ensemble as a function of the number of analysis steps taken. The panels A, B, C, and D show the evolution of the longitude, width, speed, and transit time distributions, respectively. In each panel, the grey lines mark each of the 50 ensemble members, while the blue, orange, and green lines mark the $10^{th}$ percentile, median, and $90^{th}$ percentiles. The red dashed line marks the true value of each parameter.}
\label{fig:sir_huxt_example}
\end{figure}

Figure \ref{fig:sir_huxt_example} presents an example of a SIR-HUXt analysis, showing how the ensemble evolves as a function of the number of analysis steps. These data are from a single SIR-HUXt analysis from the 100 realisations in the OSSE experiment. In this instance, the initial guess of the CME longitude, width and speed was $-4^{\circ}$, $37^{\circ}$, and $477 km~s^{-1}$, around which the initial ensemble (analysis step 0) was formed. For the longitude, speed, and transit time, the distributions evolve significantly over the analysis steps, both moving towards and reducing in spread around the true value. In this example the width distribution is less strongly impacted by the SIR analysis, drifting slightly whilst maintaining a similar spread.

Considering panel D, the uncertainty in the CME transit time (or correspondingly, arrival time), is significantly reduced from $5.3$~hours in the initial ensemble to $0.9$~hours after the SIR-HUXt analysis. These transit time errors are smaller than observed transit time errors and this is primarily due to idealised scenario used in our experiment, which does not yet include ambient solar wind structure.  

We stress that this is only one example, and that alternative behaviours are observed. Additionally, it is not "wrong" or a failure of the SIR scheme that the width distribution does not change much during the analysis. Depending on the initial estimate of the CME parameters, and the uncertainty on the observations, it is possible that the distribution of any particular parameter need not evolve significantly. 

\subsubsection{Aggregated SIR-HUXt OSSE results}

\begin{figure}
\includegraphics[width=\textwidth]{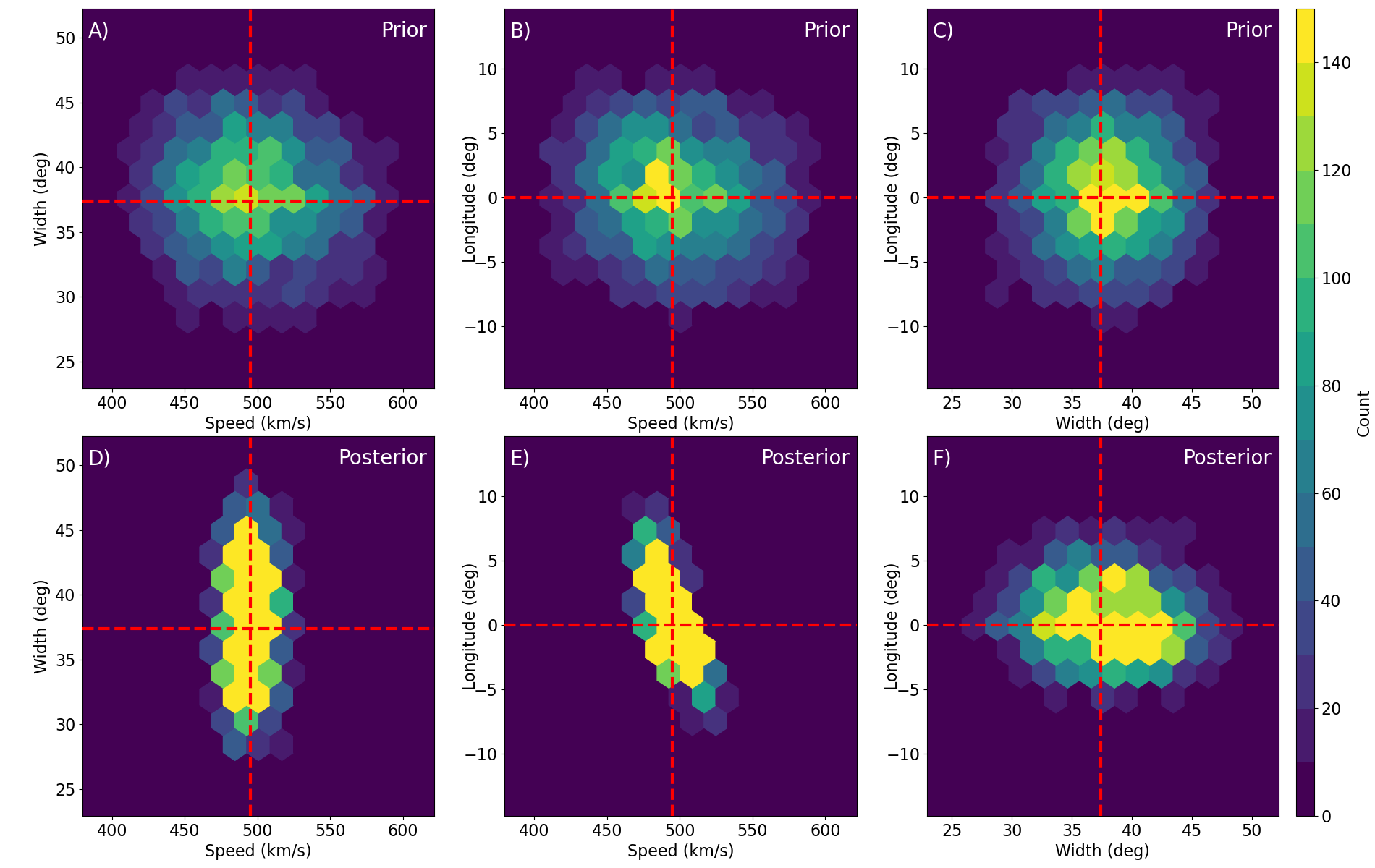}
\caption{These panels show 2-D histograms of the prior and posterior distributions of the CME parameters for the Slow-Narrow CME scenario, for an L5 Observer. The prior distributions are shown along the top row, with the posterior distributions on the bottom row. The red dashed lines mark the true parameter values.}
\label{fig:sir_huxt_aggregated}
\end{figure}

Figure \ref{fig:sir_huxt_aggregated} compares the prior and posterior distributions of the CME parameters, aggregated over all of the OSSE experiments. These are presented as three 2-D histograms, showing the joint distributions of the CME speed and width, speed and longitude, and width and longitude. The top row shows the prior distributions, whilst the bottom row shows the posterior distributions. The red dashed lines mark the true parameter values. As each SIR-HUXt run uses a 50 member ensemble, and there are 100 realisations in the OSSE, there are 5000 samples in each distribution. 

The prior distributions are relatively uniform, as expected from the perturbation function used to generate the initial ensembles. The posterior distributions have significantly different structure to the priors. Panels D and E show that the speed distribution has been strongly constrained around the true value, with the standard deviation reducing from $40~km~s^{-1}$ to $11~km~s^{-1}$. The standard deviation in width distribution is approximately the same for the prior and posterior, being $4.1^{\circ}$ and $4.3^{\circ}$. The standard deviation of the posterior longitude distribution is reduced relative to the prior, decreasing from $4.1^{\circ}$ to $3.0^{\circ}$. Panel D also shows the emergence of a correlation between the posterior distributions of speed and longitude. This is not surprising, as the time-elongation profiles of CME flanks have a well known degeneracy relating to the speed, width and longitude. Similar time-elongation profiles can be generated by CMEs travelling at different angles relative to the plane-of-sky with different speeds. This degeneracy can lead to such correlations when trying to find a combination of CME parameters that best reflects a time-elongation measurement. Consequently, we expect this to be a "feature" of assimilating CME time-elongation profiles from one observer. We note that it is possible that assimilating HI data from more than one observer, or, assimilating the HI image intensities rather than only a time-elongation profile, might break these degeneracies, and these objectives should be a priority for future investigation. Nonetheless, it is clear that the posterior CME states are typically closer to the true CME state, even if this is dominated by the evolution of the CME speed distribution.

\subsubsection{Ensemble mean SIR-HUXt OSSE results}

\begin{figure}
\includegraphics[width=\textwidth]{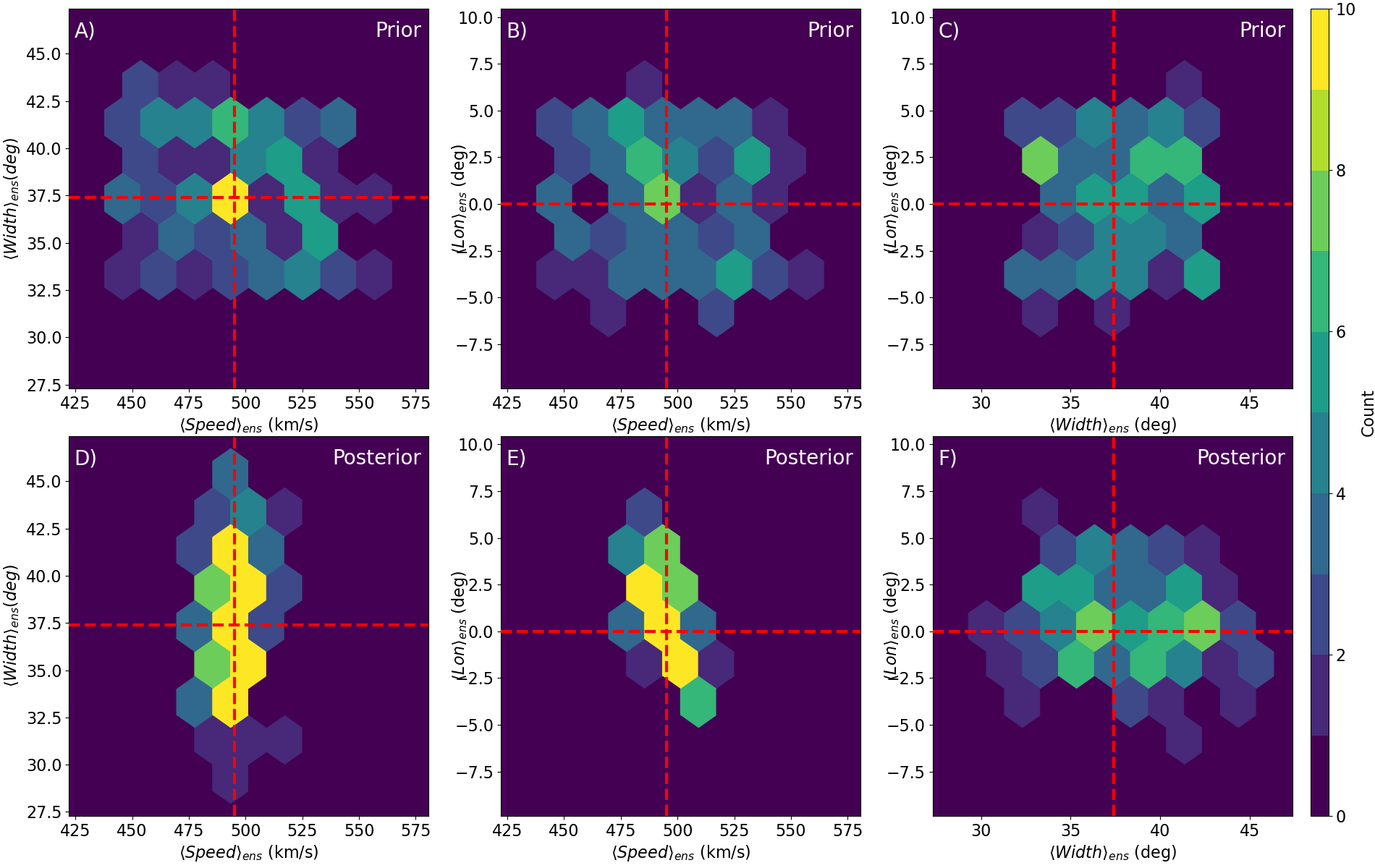}
\caption{These panels show 2-D histograms of the mean CME parameters for each SIR-HUXt realisation of the OSSE experiment. The format is the same as Figure \ref{fig:sir_huxt_aggregated}.}
\label{fig:sir_huxt_mean}
\end{figure}

It is also instructive to compare the means of the prior and posterior distributions for each realisation of the OSSE experiment. Figure \ref{fig:sir_huxt_mean} shows these data, using the same format as Figure \ref{fig:sir_huxt_aggregated}. Each histogram contains 100 samples from the 100 OSSE experiments, with each 50-member ensemble reduced to its mean value. We observe the same behaviour in these distributions as was observed for the aggregated SIR-HUXt analyses in Figure \ref{fig:sir_huxt_aggregated}. The CME speeds are strongly constrained around the true value, with the standard deviation reducing from from $28~km~s^{-1}$ to $8~km~s^{-1}$. There are only small changes between the prior and posterior distributions of the CME width and longitude.. The spread of the distribution of CME widths increases slightly with a prior standard deviation of $2.9^{\circ}$ and posterior standard deviation of $3.4^{\circ}$. Conversely, the spread of the CME longitudes decreases slightly, with a prior standard deviation of $2.9^{\circ}$ and posterior standard deviation of $2.2^{\circ}$.  

\subsubsection{Assesment of ensemble represntivity with Rank-Histograms}

It is also important to assess the representivity of the SIR-HUXt ensembles. If the SIR-HUXt ensembles were perfectly calibrated, then each ensemble member and the truth state would be independent samples from the same underlying probability distribution. A rank histogram is a graphical means of assessing this \cite{talagrand1997}. To construct the rank histogram we rank the true system state in each SIR-HUXt realisation, and plot a histogram of these data. If the truth state and ensemble members are independent samples from the same probability distribution, then the rank histogram would be uniform, to within the limits of sampling variability. However, deviations from uniformity can diagnose miscalibrations in the ensemble. For example, if the ensemble is over or under dispersed, the rank-histogram takes a U or inverted-U shape, or if the ensemble is biased the rank histogram can be asymmetric \cite{wilks2019}.

\begin{figure}
\includegraphics[width=\textwidth]{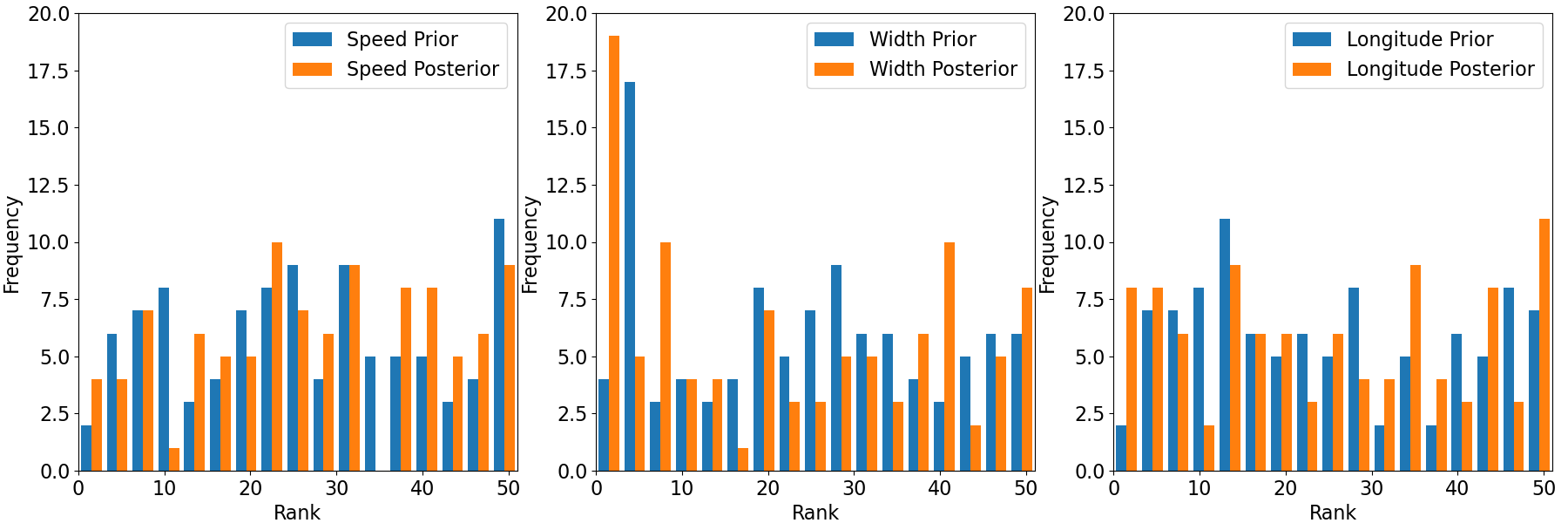}
\caption{Rank histograms of the prior and posterior distributions of the CME speed, width, and longitude}
\label{fig:sir_huxt_rank_hist}
\end{figure}

Figure \ref{fig:sir_huxt_rank_hist} presents the rank histograms for the prior and posterior distributions of CME speed, width, and longitude. As the prior distributions are generated by uniform perturbations to the CME scenario, it follows that the rank-histogram of the prior distributions are also uniform, to within the limits of sampling variability. It is clear that the posterior distributions show a similar level of uniformity, which is one indicator that the SIR-HUXt ensembles are reasonably well calibrated. 

\subsubsection{SIR-HUXt impact on CME transit time and arrival speed distributions}

\begin{figure}
\includegraphics[width=\textwidth]{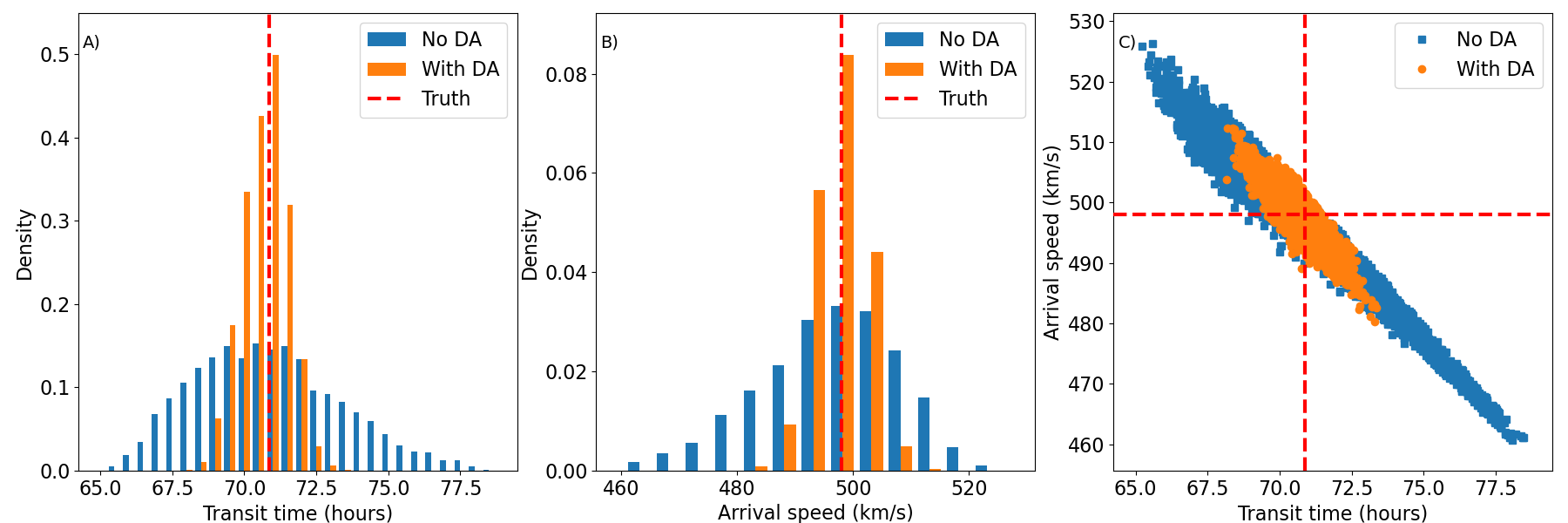}
\caption{Panels A and B show histograms of the prior and posterior distributions the CME transit time and arrival speed, respectively. Panel C presents a scatter plot of the CME transit time and arrival speed. The true CME transit time and arrival speed are shown by the red dashed lines.}
\label{fig:sir_huxt_arrival}
\end{figure}

Finally, we consider the impact of SIR-HUXt on the distributions of CME transit time and arrival speed at Earth. Figure \ref{fig:sir_huxt_arrival} presents these data. Panel A and B show histograms of the prior and posterior distributions of CME transit time and arrival speed at Earth. Panel C shows a scatter plot of the CME transit time versus the CME arrival speed for both the prior and posterior distributions. For this scenario, the true CME transit time and arrival speed were $70.8~h$, and $498~km~s^{-1}$. Considering panel A, the prior distribution has a larger spread around the true transit time than the posterior distribution. The standard deviation of the prior and posterior distributions are $2.5~h$ and $0.8~h$, respectively. Therefore the SIR-HUXt analysis results in a $69\%$ reduction in the CME transit time standard deviation. Panel B shows a similar results for prior and posterior distributions of the CME arrival speed; the posterior distribution is less spread around the true value than the prior, with standard deviations of $11~km~s^{-1}$ and $4~km~s^{-1}$, respectively, a $63\%$ reduction in the arrival speed standard deviation. Panel C shows that, as expected for this scenario without ambient solar wind structure, there is a clear anti-correlation between CME transit time and arrival speed at Earth. This correlation is present for both the joint-posterior and joint-prior distributions. However, consistent with panels A and B, the joint-posterior distribution is closer to the true CME transit time and arrival speed for this scenario. In this way, we consider this evidence that SIR-HUXt has significant potential for improving CME transit time and arrival speed simulations over simple HUXt ensembles. 

\subsection{SIR-HUXt performance with Observer longitude}

\begin{figure}
\includegraphics[width=\textwidth]{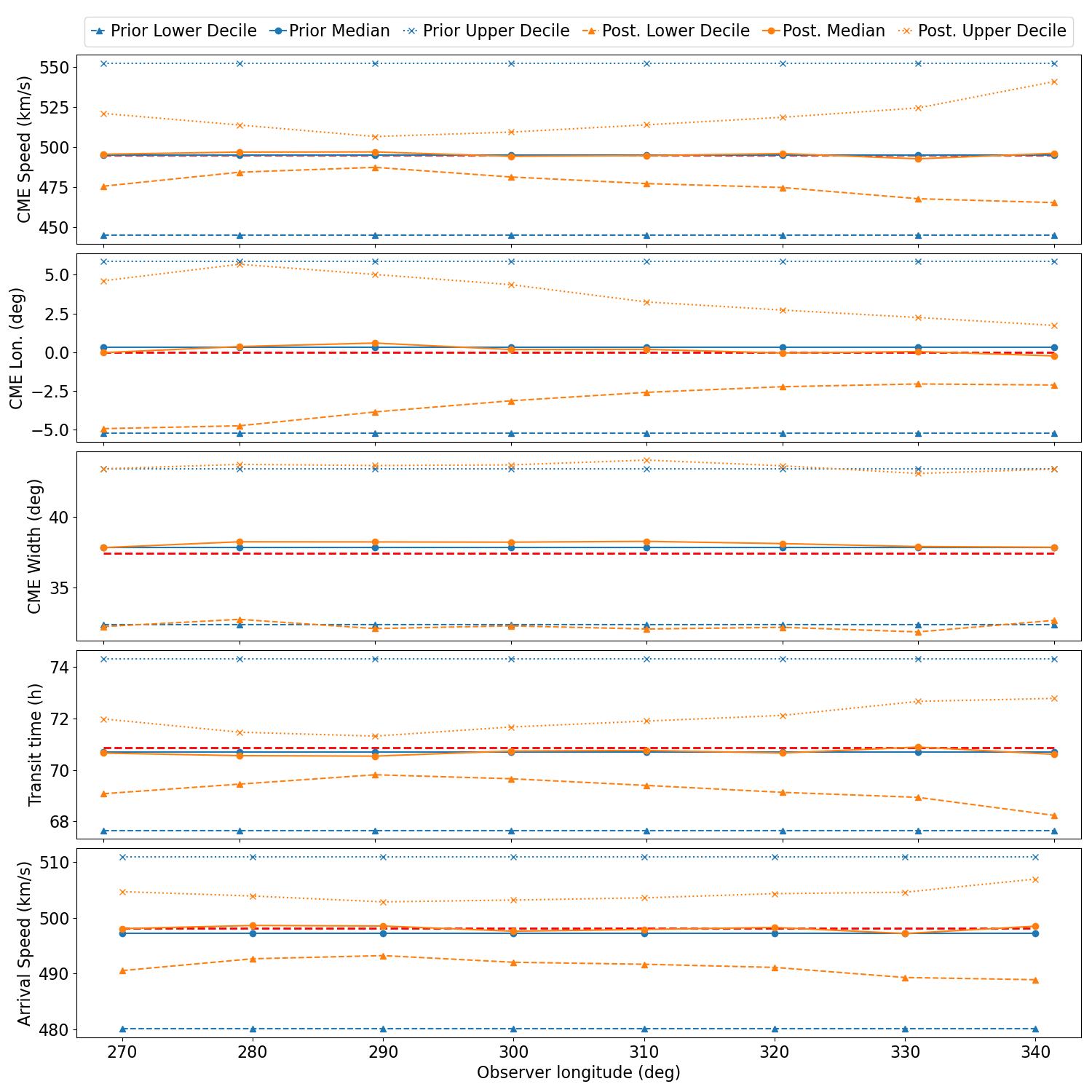}
\caption{These panels show the evolution of prior and posterior distributions of the Cone CME speed, longitude, width, transit time, and arrival speed, as a function of observer longitude. These distributions are summarised by their lower decile (triangles), median (circles) and upper decile (crosses), with the priors and posteriors colored blue and orange, respectively. The true parameter values are shown by the red dashed line}
\label{fig:params_vs_lon}
\end{figure}

We now consider how the posterior distributions of the Cone CME parameters and the transit time and arrival speed at Earth vary as a function of observer longitude. These data are presented in Figure \ref{fig:params_vs_lon}, where the prior and posterior distributions of each parameter are summarised by their lower decile, median, and upper decile. The true parameter values are shown by the red dashed line. 

The prior distributions show no variation with observer longitude, because the same set of randomly generated initial ensembles are used for the SIR-HUXt OSSE experiments at each longitude, so as to enable a fair comparison between longitudes. For each parameter, the median of the prior and posterior distributions are very similar and close to the true parameter values, indicating no significant bias between the prior and posterior distributions with the true parameter values. 

There are, however, systematic changes in the the spread of the SIR-HUXt posterior parameters as a function of observer longitude, where we define spread to be the difference between the lower and upper deciles. For the initial Cone CME speed, the posterior spread is less than the prior spread at all observing longitudes, and it also shows a local minimum at $290^{\circ}$ longitude. This indicates that the SIR-HUXt posteriors provide a tighter constraint on the CME speed from all observer longitudes, but the tightest constrain comes from assimilating time-elongation profiles from observers close to the L5 region. We think this behaviour is driven by the fact that for an observer in this region, with a fully Earth directed CME, the time-elongation profile of the flank corresponds closely to the CME apex (see Figure \ref{fig:huxt_cme_scenario} panel B), which minimises the degeneracy between the CME speed, longitude and width. However, this is not the case for the CME longitude, where the spread of the distribution continues to increase as the observer moves from $340^{\circ}$ to $290^{\circ}$. This suggests that SIR-HUXt is better able to constrain the CME source longitude from Observations nearer Earth. There seem to be no significant differences between the prior and posterior distributions of the CME width, suggesting that for this particular scenario SIR-HUXt does not have a significant impact on the CME width estimation. Further experiments are required to determine whether this behaviour is general, or is specific to this particular CME scenario. Both the CME transit time and arrival speed show behaviour that mirrors that of the CME speed, with the spread being less than the prior for all observing longitudes, and showing a local minimum at $290^{\circ}$. This is unsurprising, given that for this CME scenario, with uniform ambient solar wind, we expect the transit time and arrival speed to be primarily determined by the Cone CME initial speed.

\section{Conclusions}
\label{sec:conclusions}

In this work we have presented the development of SIR-HUXt, a particle filter data assimilation (DA) scheme for constraining the HUXt solar wind model. SIR-HUXt assimilates time-elongation profiles of a CMEs flank, which is a data product routinely derived from heliospheric imager data, such as that returned by STEREO-HI, Parker Solar Probe's WISPR. 

The motivation for pursuing the development of SIR-HUXt is that, at present, there is significant uncertainty in the initial and boundary conditions of the solar wind numerical models that are used for both scientific and forecasting purposes. These uncertainties limit both the scientific inferences and forecast skill we can extract from solar wind simulations. DA techniques present a pathway to reduce these uncertainties, providing a framework for combining simulations with observations to return an optimal estimate of the state of the system. HUXt is well suited to the development of DA schemes due to it's simplicity and low computational expense. This latter point is particularly important for the development of this Sequential Importance Resample particle filter, which requires a large ensemble of simulations to be run ($\approx~500$ per SIR-HUXt analysis); this would be very computationally expensive for full 3D MHD solar wind models.

In it's current form, SIR-HUXt adjusts the Cone CME parameters in response to assimilating the time-elongation profile of an observed CMEs flank. We used Observing System Simulation Experiments (OSSEs) to provide an initial proof-of-concept test of the SIR-HUXt algorithm. These experiments test the ability of SIR-HUXt to recover a known synthetic truth state, by assimilating pseudo-observations of the known truth state. In these experiments, our truth state was a simple scenario of an Earth directed CME with the observed median speed and width, propagating through a uniform ambient solar wind. These OSSEs showed that SIR-HUXt is effective at constraining the CME state, primarily by adjusting the CME speed. These experiments suggested SIR-HUXt was less effective at constraining the CME longitude and width. Nonetheless, by improving the constraint of the initial CME state, SIR-HUXt also returns improved estimates of the CME transit time to Earth, and the CMEs arrival speed, which are critical parameters for space weather predictions. The reliability of the SIR-HUXt ensembles was assessed through rank-histogram plots, through which we conclude that the SIR-HUXt ensembles are reasonably well calibrated, with no clear indications of under or over dispersion, or bias.

The OSSEs also revealed that the location of the observer relative to the CME has a significant impact on the ability of SIR-HUXt to constrain the CMEs parameters. Observers near the L5 location provided the best constraints on the CME speed, transit time and arrival speed. Whilst the SIR-HUXt constraints on the CME longitude grew weaker as separation between the observers longitude and CME apex longitude increased. This is a potentially significant result relating to likely performance of schemes like SIR-HUXt for space weather forecasting with the operational heliospehric imager data that will be returned by ESA's Vigil mission.

We note that it is now well established that both the CME initial conditions and the ambient solar wind structure both play important roles in determining the simulated CME evolution and, critically, the forecast arrival time at Earth. Indeed, \cite{riley2021} investigated the sources of uncertainty in CME arrival time predictions are concluded that both the ambient solar wind structure and CME parameters introduce similar magnitudes of uncertainty. Our work has so far only considered uncertainty in the CME parameters, and does not yet consider uncertainty in the ambient solar wind structure, although we have plans to tackle this issue in future work. 

Following this study, our next two objectives are to test SIR-HUXt with OSSEs using a wider range of CME scenarios, and to test SIR-HUXt with actual time-elongation profiles extracted from the STEREO-HI data. 

\section{Open Research}
The software to generate all of the simulation data used in this study are available at: \url{https://github.com/LukeBarnard/SIR_HUXt}

\section{Acknowledgements}
This research made use of Astropy (\url{http://www.astropy.org}), a community-developed core Python package for Astronomy \cite{robitaille2013, price-whelan2018}.

This research used version 4.0.0 \cite{mumford2022} of the SunPy open source software package \cite{barnes2020b}.

Figures for this article were made with version 3.3.4 of Matplotlib \cite{caswell2021, hunter2007}

This work was part-funded by Science and Technology Facilities Council (STFC) grant numbers ST/R000921/1 and ST/V000497/1, and NERC grant number NE/S010033/1.

\end{document}